\documentstyle[12pt]{article}
\makeatletter\newdimen\normalarrayskip              
lines
\newdimen\minarrayskip                 
\normalarrayskip\baselineskip
\minarrayskip\jot
\newif\ifold             \oldtrue            \def\new{\oldfalse}
\def\arraymode{\ifold\relax\else\displaystyle\fi} 
enrties
\def\eqnumphantom{\phantom{(\theequation)}}     
eqnarray
\def\@arrayskip{\ifold\baselineskip\z@\lineskip\z@
     \else
     \baselineskip\minarrayskip\lineskip2\minarrayskip\fi}
\def\@arrayclassz{\ifcase \@lastchclass \@acolampacol \or
\@ampacol \or \or \or \@addamp \or
   \@acolampacol \or \@firstampfalse \@acol \fi
\edef\@preamble{\@preamble
  \ifcase \@chnum
     \hfil$\relax\arraymode\@sharp$\hfil
     \or $\relax\arraymode\@sharp$\hfil
     \or \hfil$\relax\arraymode\@sharp$\fi}}
\def\@array[#1]#2{\setbox\@arstrutbox=\hbox{\vrule
     height\arraystretch \ht\strutbox
     depth\arraystretch \dp\strutbox
     width\z@}\@mkpream{#2}\edef\@preamble{\halign
\noexpand\@halignto
\bgroup \tabskip\z@ \@arstrut \@preamble \tabskip\z@ \cr}%
\let\@startpbox\@@startpbox \let\@endpbox\@@endpbox
  \if #1t\vtop \else \if#1b\vbox \else \vcenter \fi\fi
  \bgroup \let\par\relax
  \let\@sharp##\let\protect\relax
  \@arrayskip\@preamble}
\def\eqnarray{\stepcounter{equation}%
              \let\@currentlabel=\theequation
              \global\@eqnswtrue
              \global\@eqcnt\z@
              \tabskip\@centering
              \let\\=\@eqncr
              $$%
 \halign to \displaywidth\bgroup
    \eqnumphantom\@eqnsel\hskip\@centering
    $\displaystyle \tabskip\z@ {##}$%
    &\global\@eqcnt\@ne \hskip 2\arraycolsep
         \hfil$\arraymode{##}$\hfil
    &\global\@eqcnt\tw@ \hskip 2\arraycolsep
         $\displaystyle\tabskip\z@{##}$\hfil
         \tabskip\@centering
    &{##}\tabskip\z@\cr}
\makeatother

\def\beq{\begin{equation}}
\def\eeq{\end{equation}}
\def\bea{\begin{eqnarray}}
\def\eea{\end{eqnarray}}

\newcommand{\m}[1]{\makebox[.40in]{#1}}
\newcommand{\ml}[1]{\makebox[.40in][l]{#1}}
\newcommand{\mr}[1]{\makebox[.40in][r]{#1}}
\newcommand{\n}[1]{\makebox[.30in]{#1}}

\def\s{\stackrel} \def\ts{\textstyle} \def\st{\scriptstyle}

\begin{document}
\begin{titlepage}
\begin{center}
\hfill UUITP\\
\hfill 9/1993\\
 \hfill hepth@xxx/93\#\#
\begin{flushright}{March 1993}\end{flushright}
\vspace{0.1in}{\Large\bf Ward identities and W-constraints \\
in Generalized Kontsevich Model }\\[.4in]
{\large  A.Mikhailov}\footnote{On leave from The Moscow State
University, Physical Department, Moscow, Lenin hills, Russia}\\
\bigskip {\it
Institute of Theoretical Physics \\
Uppsala University\\
Box 803 S-75108\\
Uppsala, Sweden\\
mika@titan.teorfys.uu.se}

\end{center}
\bigskip \bigskip

\begin{abstract}
The connection is obtained between Ward identities and
W-con\-straints in
Generalized Kontsevich Model with the potential $X^4/4$.
We show that Ward identities include W-constraints (and do not
include any other
constraints) for this potential and make some observations in favour
of
 the same connection
for the model with the potential of the form $X^{(K+1)}/(K+1)$ for
any $K\geq 2$.
\end{abstract}

\end{titlepage}
\section{Introduction}\label{Introduction}

Matrix models play the very important role in string theory. They
allow to
investigate essentially nonperturbative aspects of $2d$ quantum
gravity. One of
the most interesting properties of these models is their connection
with the
classical integrable equations: partition functions of these models
(or some
simple functions of them) are
$\tau$-functions \cite{Dou90}. What is specific in the
$\tau$-functions which can be
obtained from matrix integrals is that they satisfy string equation,
or
${\cal L}_{-1}$-constraint. It was proved in \cite{FKN91b} that if
$\tau$-function
satisfy this constraint and does not depends on $T_{Kn}$ then it
satisfy the
$W_K$-algebra vacuum conditions. This is the property of $\tau$ which
can be proved
by analyzing the consequences of ${\cal L}_{-1}$-constraint as the
constraint
imposed on the point of the Grassmanian manifold the $\tau$-function
is
associated with. But we thing that it is more natural to extract
these
{\cal W}-constraints directly from the matrix integral, as some
purely
arithmetic fact.

 The conventional matrix model approach to 2d gravity is based on the
Hermitean
multimatrix model with the partition function
    \begin{equation}  \Gamma ^{K}=\prod_{i=1}^{K} \int
dM_{i}\exp(\sum t_{n}trM_{i}^{n}+
    trM_{i}M_{i+1})
    \label{statsum of Hermitean multimatrix model}
    \end{equation}

 Such models are called "discrete", because we need a
sophisticated double scaling limit to obtain continuous theories from
them.
These continuous theories are minimal models coupled to $2d$ gravity
\cite{Dou90,Kaz89,BK90,DS90,GM90c}.
 There exists however an alternative way based on the Generalized
Kontsevich Model.
 This is a one matrix model in the external field with the partition
function:
 \begin{equation} Z_{N}^{\{{\cal V}\}}=\frac{\int e^{-U(M,
Y)}dY}{\int e^{-U_{2}(M, Y)}dY}
                \label{statsum of Generalized Kontsevich Model}
 \end{equation}
 where  \( U(M, Y)=tr[{\cal V}(M+Y)-{\cal V}(M)-{\cal V}'(M)Y] \) \ \
\ \    and \\
\( U_{2}(M, Y)=\lim_{\epsilon \rightarrow \infty} \frac{1}{\epsilon
^{2}} U(M, \epsilon Y) \).

   It was argued in paper \cite{KMMMZ91b} that for specially adjusted
potential
   ${\cal V}(X)$  it    describes the double scaling limit of any
multimatrix model, namely:
\begin{equation}
Z_{\infty}^{\{K\}}[M]=\sqrt{\lim_{d.s.}\Gamma^{\{K\}}}
                \label{connection between GKM and HMMM}
\end{equation}
   where superscription $\{K\}$ means that \({\cal
V}=\frac{X^{K+1}}{K+1}\). An
   important argument to perform such identification is the existence
of
   ${\cal W}$-constraints:
\begin{equation} {\cal W}_{Kn}^{\{K\}}[M]=0, \; n\geq1-k
                \label{W-constraints}
\end{equation}

   It was shown in \cite{FKN91a,DVV91a}, that the RHS of
(\ref{connection
   between GKM and HMMM}) satisfies the same constraints. Here
   $Z_{N}^{\{K\}}[M]$ is expressed through times
$T_{m}=\frac{1}{m}trM^{-m}$,
    and ${\cal W}$-con\-straints can    be expressed through free
fields $mT_{m}$,
    $\partial/\partial T_{m}$ \cite{FKN91a,FKN91b}, for example
($K=3$):

 \begin{equation}\new\begin{array}{c}
 {\cal W}_{3n}^{\{3\}}=3\sum_{k, l\geq1}k\hat{T_{k}}l\hat{T_{l}}
\frac{\partial}{\partial
T_{k+l+3n}}+3\sum_{k\geq1}k\hat{T_{k}}\sum_{a+b=k+3n}
\frac{\partial^2}{\partial T_{a}\partial
T_{b}}+\\+\sum_{a+b+c=3n}\frac{\partial^{3}}
{\partial T_{a}\partial T_{b}\partial T_{c}}+\\+
\sum_{a+b+c=-3n}a\hat{T_{a}}b\hat{T_{b}}c\hat{T_{c}}; a, b, c>0, a,
b, c, k, l\neq0
\bmod 3
\end{array}
                 \label{W3 operator}
 \end{equation}

\begin{equation}\new\begin{array}{c}
{\cal W}_{3n}^{\{2\}}=\frac{1}{3}\sum_{k}k\hat{T_{k}}\frac{\partial}
{\partial T_{k+3n}}+\frac{1}{6}\sum_{{a+b=3n}\atop{a, b>0}}
\frac{\partial^{2}}{\partial T_{a}\partial T_{b}}+\\+
\frac{1}{6}\sum_{{a+b=-3n}\atop{a,b>0}}a\hat{T}_{a}b\hat{T}_{b}
+\frac{1}{9}\delta_{n,0}
\end{array}
                \label{W2 operator}
 \end{equation}

   Here
  \begin{equation} (\hat{T})_n=T_{n}-\frac{3}{4}\delta_{n,4}
                \label{timeshift}
  \end{equation}

   It was suggested in \cite{KMMMZ91b} that these ${\cal W}$
-constraints are connected with
   Ward identities in the following way. Consider a matrix integral
\begin{equation} {\cal F}^{\{K\}}[\Lambda]=\int
dXe^{-tr\frac{X^{K+1}}{K+1}+tr\Lambda X}
                \label{matrix Airy function}
\end{equation}
   It can be expressed through  partition function of GKM by means of
a formula:
\begin{equation}
{\cal F}^{\{K\}}[\Lambda]=g_{K}[\Lambda]Z_{N}^{\{K\}}[{\cal M}]
                \label{connection between GKM and matrix Airy
function}
\end{equation}
   where \(\Lambda={\cal M}^{K}\) and
\begin{equation}\new\begin{array}{c}
g_{K}[\Lambda]=\frac{\Delta(\Lambda^{1/K})}{\Delta(\Lambda)}
\prod_{i}[\lambda_{i}^{-\frac{K-1}{2K}}
e^{\alpha\frac{K}{K+1}\lambda_{i}^{1+1/K}}]=\\=
det\frac{1}{\sqrt{\sum_{p=0}^{K-1}\Lambda^{\frac{K-1-p}{K}}\otimes\Lam
bda^{\frac{p}{K}}}}
\prod_{i}e^{\alpha\frac{K}{K+1}\lambda_{i}^{1+1/K}}
\end{array}
                 \label{prefactor}
\end{equation}
   where $\alpha=1$ is a parameter, which is convenient from the
point of view  of
   "dimension". $ Z_{N}^{\{K\}}$ satisfies obvious Ward identities
\begin{equation}
tr\left(\epsilon\left[\frac{\partial^{K}}{\partial\Lambda_{tr}^{K}}-\a
lpha^{K}
\Lambda\right]\right)g_{K}[\Lambda]Z^{\{K\}}(T_{n})=0
                \label{Ward identities}
 \end{equation}
 where $\epsilon$  is arbitrary matrix. The suggestion is that after
substitution
(\ref{connection between GKM and matrix Airy function}) in (\ref{Ward
identities})
 where $Z$ is expressed through times, this equation will be some
   combination of ${\cal W}$-constraints. This suggestion was proved
in
   \cite{MMM92a} for $K=2$ (ordinary Kontsevich model,
   constraints form the Virasoro algebra).
   In this paper we prove a formula
\begin{equation}
\new
\begin{array}{c}
\frac{1}{g_{3}[\Lambda]}tr \: \epsilon\left(\frac{\partial^{3}}
{\partial^{3}\Lambda_{tr}^{3}}-\alpha^{3}\Lambda\right)g_{3}[\Lambda]Z
^{\{3\}}(T_{n})=\\=
-\frac{1}{9} tr\{\epsilon{\cal M}^{-9}[\sum_{n\geq-2}\frac{1}{3}
{\cal M}^{-3n}{\cal W}_{3n}^{(3)}+\\
 +3\sum_{n\in Z }{\cal
M}^{-3n+1}\sum_{q\geq0}(3q+1)\theta(n+q+1)\hat{T}_{3q+1}
 {\cal W}_{3n+3q}^{(2)}+\\+ 3\sum_{n\in Z}{\cal
M}^{-3n+2}\sum_{q\geq0}(3q+2)
 \theta(n+q+1)\hat{T}_{3q+2}{\cal W}_{3n+3q}^{(2)}-\\-
 6\sum_{n\geq3}{\cal M}^{3n}\sum_{q\geq0}3q\theta(-n+q+1)\hat{T}_{3q}
 {\cal W}_{(-3n+3q)}^{(2)}+\\+
 3\sum_{n\geq-1}{\cal M}^{-3n-1}\sum_{{a\geq0, b\geq-1}\atop{a+b=n}}
 \frac{\partial}{\partial T_{3a+1}}{\cal W}_{3b}^{(2)}+\\+
 3\sum_{n\geq-1}{\cal M}^{-3n-2}\sum_{{a\geq0,b\geq-1}\atop{a+b=n}}
 \frac{\partial}{\partial T_{3a+2}}{\cal
W}_{3b}^{(2)}]\}Z^{\{3\}}(T_{n})
                \label{MAINFORMULA}
\end{array}
\end{equation}
which establishes the desired relation in $K=3$ (Boussinesq) case and
obtain some
reasons why formulae of such a type hold for every $K$. Here
$\theta(m)$ is equal to 1
if $m\geq0$ and 0 otherwise. The sums in this expression should be
understood in the Borel sense.
What we obtained is a system of equations of the type:
 \begin{equation}\new\begin{array}{c}
 \sum f_{n}(T)\mu_{1}^{n}=0\\
 \cdots                    \\
 \sum f_{n}(T)\mu_{N}^{n}=0
 \end{array}\label{bore1}\end{equation}
 Due to the symmetry under permutations
 of eigenvalues we have actually only one equation, for example the
first one.
 Now let us vary $\mu_{1}$ in open domain in this equation. If we
have
 sufficiently many Miwa variables, we can vary other of them in such
a way, that
 coefficients $f_{n}$ will change only by a small amount. This amount
can be
 done as small as we want if we let $N\rightarrow\infty$. So in this
 limit we have a series in $\mu_{1}$ with constant coefficients,
which is identically
 equal to zero. So the coefficients are equal to zero. Since the
coefficients
 (expressed through times) do not formally depend on $N$, they are
{\em equal
 to zero} for finite $N$.
 An important
property of the formula is that terms with  ${\cal M}^{3n},n\geq3$,
are linearly independent
combinations of the ${\cal W}$-operators, and,
therefore, we infer that $Z$ is annihilated by
them. Recently there was a paper \cite{IZ92},
in which a piece of this formula was obtained.
This piece contains some  ${\cal W}$-operators, and all others can be
obtained from them by commutating. Our formula
proves that all the ${\cal W}$-operators are in fact included into
Ward identities, and that no any other constraints imposed on  $Z$
can be inferred from them.

The text is organized as follows. In sections 2,3 and 5 we calculate
the terms in the LHS of the Ward identities, which we need to prove
it's equivalence to the $W$-constraints. In section 4 we describe
some mathematical technique which we think is useful in the
calculations of such a type. In Appendix we make some observations
about the general $K$ case.
 Fix notations: $\mu_{i}$ are eigenvalues of  ${\cal M}$,
 $\nu_{i}=\mu_{i}^{-1}$, $T_{n}=\frac{1}{n}\sum_{i}\mu_{i}^{-n}$,
 $t_{n}=nT_{n}$.

 \section{Terms with $\frac{\partial^{3}Z}{\partial
T^{3}}$.}\label{threeder}
 Using the explicit expressions for times, we obtain:
 \(\partial T_{m}/\partial\lambda_{i}=-\frac{1}{3}\mu_{i}^{-(m+3)}\),
so the term
 of interest is
\begin{equation}
-\frac{1}{27}\sum_{m,n,l}\mu_{i}^{-(m+n+l+9)}\frac{\partial^{3}Z}
{\partial T_{m}\partial T_{n}\partial T_{l}}
             \label{term with three derivatives}
\end{equation}

 \section{Terms with \(\frac{\partial^{2}Z}{\partial
T^{2}}\).}\label{twoder}
 They arise from the terms:
\begin{equation}\new\begin{array}{c}
           \sum_{m,n}\left(\sum_{j}g\frac{\partial^{2}T_{n}}
{\partial\Lambda_{ij}\partial\Lambda_{ji}}\frac{\partial
T_{m}}{\partial\lambda_{j}}
\frac{\partial^{2}Z}{\partial T_{m}\partial T_{n}}+
                            2\sum_{j}g\frac{\partial^{2}T_{n}}
{\partial\Lambda_{ij}\partial\Lambda_{ji}}\frac{\partial
T_{m}}{\partial\lambda_{i}}
\frac{\partial^{2}Z}{\partial T_{m}\partial T_{n}}+\right.\\+\left.
                            3\frac{\partial g}{\partial\lambda_{i}}
\frac{\partial T_{m}}{\partial\lambda_{i}}\frac{\partial
T_{n}}{\partial\lambda_{i}}
\frac{\partial^{2}Z}{\partial T_{m}\partial T_{n}}\right)
 \end{array}
 \end{equation}
               \label{part containing three derivatives}

 The second derivatives of T can be calculated from the quantum
mechanical formula for
the shift of eigenvalues in the second order of perturbation theory
(see the section
{}~\ref{Mathematical technique}). Now consider the part which does not
contain $\alpha$.
Adding the same expression with $m$ and $n$ exchanged one obtains :
 \begin{equation}\new\begin{array}{c}
 -\frac{1}{18}\frac{1}{1+\nu_{j}/\nu_{i}+\nu_{j}^{2}/\nu_{i}^{2}}
\sum_{j,m,n}\{\sum_{p=0}^{n+2}(\nu_{i}^{n+3-p}\nu_{j}^{6+m+p}
+2\nu_{i}^{m+n+6-p}\nu_{j}^{3+p})+\\+\sum_{p=0}^{m+2}(\nu_{i}^{m+3-p}\
nu_{j}^{6+n+p}
+2\nu_{i}^{m+n+6-p}\nu_{j}^{3+p})
+4\nu_{i}^{m+n+7}\nu_{j}^{2}+2\nu_{i}^{m+n+8}\nu_{j}\}
\end{array}
\end{equation}
 Here we must consider separately cases when $m\mbox{\ and\ }n$
belong to different
 residue classes modulo 3, but we must consider only nonzero classes,
because $Z^{\{K\}}$
does not depend on $T_{nK}$ \cite{KMMMZ91b}. Consider the cases:

 1) $m=n=1 \bmod 3$.
\begin{equation}\new\begin{array}{c}
 2\sum_{p=0}^{(m+n+4)/3}t_{3p+2}{\cal M}^{-m-n-7+3p}
 +\sum_{p=0}^{(n+2)/3}t_{3p+1}{\cal M}^{-m-n-8+3p}+\\
 +\sum_{p=0}^{(m+2)/3}t_{3p+1}{\cal M}^{-m-n-8+3p}\equiv
 A_{11}+B_{11}+C_{11}
\end{array}\end{equation}

 2) $m=n=2 \bmod 3$
\begin{equation}\begin{array}{c}
 2\sum_{p=0}^{(m+n+5)/3}t_{3p+1}{\cal M}^{-m-n-8+3p}+
 \sum_{p=0}^{(n+1)/3}t_{3p+2}{\cal M}^{-m-n-7+3p}+\\+
 \sum_{p=0}^{(m+1)/3}t_{3p+2}{\cal M}^{-m-n-7+3p}\equiv
 A_{22}+B_{22}+C_{22}
\end{array}\end{equation}

 3) $m=1,n=2 \bmod 3$
\begin{equation}\begin{array}{c}
 -2\sum_{p=(m+n)/3+2}^{\infty}{\cal M}^{-m-n-9+3p+3}t_{3p+3}+
 2\sum_{p=0}^{(m+2)/3}{\cal M}^{-m-n-8+3p}t_{3p+1}+\\+
 2\sum_{p=0}^{(n+1)/3}{\cal M}^{-m-n-7+3p}t_{3p+2}+
 \sum_{p=1}^{\infty}{\cal
M}^{-n-6+3p}t_{m+3+3p}+\\+\sum_{p=1}^{-m-6+3p}t_{n+3+3p}\equiv\\
  A_{12}+B_{12}+C_{12}+D_{12}+E_{12}
\end{array}\end{equation}
(We omitted for brevity the factor -1/18.)

  It turns out that $B_{11}+C_{11}+\frac{1}{2}C_{12}$ is equal to the
sum of the
  second derivatives in the expression  \(6\sum_{M>6,M=1 \bmod 3}
  {\cal M}^{-M}\sum_{\ts\s{m+3q=M-9}{\st m\geq0,m=1\bmod 3}}
  \frac{\partial}{\partial T_{m}}{\cal W}_{3q}^{(2)}\),
$A_{11}+A_{22}$ is
  equal to the same sum in \(\frac{2}{3}\sum_{n\geq-2}{\cal
M}^{-3n}{\cal W}_{3n}^{(3)}
  \).$B_{22}+C_{22}+\frac{1}{2}B_{12}$ is equal to the same sum in
the expression
  \(6\sum_{\ts\s{N>6}{\st N=2\bmod3}}{\cal M}^{-N}
  \sum_{\s{n+3q=N-9,n>0}{\st n=2\bmod3\\,q\geq-1}}
  \frac{\partial}{\partial T_{n}}{\cal W}^{(2)}_{3q} \), and
\\$\frac{1}{2}(B_{12}+C_{12})
  +E_{12}+D_{12}+(m\leftrightarrow n)$---in
  \(6\sum_{\ts\s{N\neq0\bmod3,q\geq0}{\st q\neq N,0\bmod 3}}{\cal
M}^{-N}
  qT_{q}{\cal W}_{N+q-9}^{(2)}\). \\$A_{12}$ yields the second
derivatives in\\
  \( -12\sum_{\ts\s{N_{+}\geq0}{\st N_{+}=0\bmod3}}{\cal M}^{N_{+}}
  \sum_{{q\geq0}\atop{q=0\bmod3}}qT_{q}{\cal W}_{-N_{+}+q-9}^{(2)}
\).

\section{Some mathematical technique: difference de\-ri\-va\-ti\-ves,
sym\-met\-ric
polynomials, Newton diagrams.}\label{Mathematical technique}
  Here we will discuss some technique which we suppose is inherent in
calculations
 in matrix models. We intend to use it in the next section. Consider
a smooth
 function $f$ and introduce the difference derivative:
 \begin{equation}\new\begin{array}{c}
 \delta^{(N)}f(x_{0},x_{1},\ldots,x_{N})=\\
 =\frac{f(x_{0})}{(x_{0}-x_{1})\ldots(x_{0}-x_{N})}+\mbox{cycl.
\(x_{0},\ldots,x_{N}\)}
 \end{array}
 \end{equation}
 This definition can be extended to the case of some arguments of
$\delta^{(N)}f$
 coinciding by means of l'Hospital rule. Important properties of
$\delta^{(N)}f$
 are:
 \begin{eqnarray}
 \delta^{(N)}f(x_{0},\ldots,x_{N+1})=\frac{\delta^{(N)}f(x_{0},\ldots,
x_{N})-
 \delta^{(N)}f(x_{1},\ldots,x_{N+1})}{x_{0}-x_{N+1}}\\
 \delta^{N}f(x,\ldots,x)=\frac{1}{N!}f^{N}(x)
 \end{eqnarray}
 Thus, the Taylor expansion can be written:
 \begin{equation}
 f(x+a)=f(x)+\delta
f(x,x)+\ldots+\delta^{(N)}f(\underbrace{x,\ldots,x}_{N+1})
 a^{N}+o(a^{N})
 \end{equation}
 What is important for us is that this expression is valid for
functions of
 operators if we will take care of ordering, {\em viz.}:
 \begin{equation}
 f(X+A)=f(X)+\delta
f(\s{3}{X},\s{1}{X})\s{2}{A}+\ldots+\delta^{N}f(\s{2N+1}{X},
 \ldots,\s{1}{X})\s{2N}{A}\ldots\s{2}{A}+o(A^{N})
 \label{M1}\end{equation}
 where numbers above operators denote their order.
 See {\em e.g.} \cite{KarMas} for rigorous proof. Here we need to
take a derivative
 of the form:
 \begin{equation}
 \sum_{j_{1},\ldots,j_{K}}\frac{\partial^{K}}{\partial
 \Lambda_{ij_{1}}\partial\Lambda_{j_{1}j_{2}}\ldots\partial\Lambda_{j_
{K}i}}
 tr\/f(\Lambda)=:D^{K}_{i}f
 \label{M2}\end{equation}
After taking the trace of eq. (\ref{M1}) it reduces to the problem to
calculate the
derivatives of products $\Lambda\cdots\Lambda$. The answer is as
follows.
Consider the graphs, consisting of a circle with an orientation and
$K$ points
on it, two of them being marked. We will call marked points as
$P_{1}$ and $P_{2}$; every point can be connected with some other
points by dashed
lines, in such a way that ''$A$ connected with $B$ '' is the
equivalence relation.
 To each such picture we associate a term in $D_{i}^{(K)}f$ of the
form:
\begin{equation}
N(\mbox{picture})\sum_{j_{P_{1}},\ldots,j_{P_{K}}}
\delta^{(K+1)}f(\lambda_{j_{P_{1}}},\lambda_{j_{P_{1}}},\lambda_{j_{P_
{2}}},
\lambda_{j_{P_{3}}},\ldots,\lambda_{j_{P_{K}}})
\end{equation}
where summation is performed over the sets
$(j_{P_{1}},\ldots,j_{P_{K}})$,
such that \(j_{P_2}=i\) and \(j_{P_{j}}=j_{P_{k}}\), if $P_{j}$ and
$P_{K}$ are
connected. Here $N(\mbox{picture})$ is the number of ways to go
around the
circle from $P_{2}$, passing each arc between two neighbor points
exactly
once and having the possibility to jump between connected points.
 It can be easily inferred from this procedure that if
 \(\frac{\partial^{2}f}{\partial\lambda_{i}\partial\lambda_{j}}=
 \delta_{ij}\frac{\partial^{2}f}{\partial\lambda_{i}^{2}}\) then
 \begin{equation}
 \frac{\partial^{3}f}
 {\partial\Lambda_{ij}\partial\Lambda_{jk}\partial\Lambda_{ki}}=
  \sum_{jk}{\cal P}_{(ijk)}\left(\frac{\partial
f/\partial\lambda_{i}}

{(\lambda_{i}-\lambda_{j})(\lambda_{i}-\lambda_{k})}\right)+\frac{1}{2
}
  \frac{\partial^{3}f}{\partial\lambda_{i}^{3}}
  \label{Born}\end{equation}
  where ${\cal P}$ means {\em cyclic} permutations.

  Now we'll remind the concept of the Newton diagram. Consider the
polynomial of $N$
  variables with integer coefficients. We associate each monomial
  in it to a point in $R^{N}$, for which the $i$-th coordinate is
equal to the
  power of $x_{i}$ in the monomial, and define multiplicity of a
point as a
  coefficient before the monomial. If this monomial is a homogeneous
one of some
  degree $d$, then the Newton diagram will lie in the plane,
intersecting
  the coordinate axes in the points with the  coordinate $d$.

  Applying formula (\ref{Born}) to $T_{m}$, we obtain for the first
item on
  the RHS:
  \begin{equation}\new\begin{array}{c}
  \sum_{jk}{\cal P}_{ijk}\frac{\partial
T_{m}/\partial\lambda_{i}}{(\lambda_{i}
  -\lambda_{j})(\lambda_{i}-\lambda_{k})}=\\
  =\sum_{jk}\left[-\frac{(\nu_{i}\nu_{j}\nu_{k})^{3}}{3}\right]
  \frac{\left|\begin{array}{ccc}1&1&1\\
  \nu_{i}^{3}&\nu_{j}^{3}&\nu_{k}^{3}\\
  \nu_{i}^{6+m}&\nu_{j}^{6+m}&\nu_{k}^{6+m}\end{array}\right|}
  {\left|\begin{array}{ccc}1&1&1\\
  \nu_{i}^{3}&\nu_{j}^{3}&\nu_{k}^{3}\\
  \nu_{i}^{6}&\nu_{j}^{6}&\nu_{k}^{6}\end{array}\right|}
  \end{array}
  \label{permutohedron}
  \end{equation}
  Such ratio is a peculiar case of Schur functions. When $m$ is
congruent to
  0 modulo 3, it is a full symmetric polynomial
$h_{m/3}(\nu_{i}^{3},\nu_{j}^{3},
  \nu_{k}^{3})$, so, for $m$ noncongruent to 0 it is something like
the
  ''full symmetric polynomial of fractional degree'', but this is not
a
  polynomial but a rational function. It's denominator is
  \( (\nu_i^2+\nu_i\nu_j+\nu_j^2)(\nu_i^2+\nu_i\nu_k+\nu_k^2)
  (\nu_j^2+\nu_j\nu_k+\nu_k^2) \). The numerator is the polynomial,
whose
  Newton diagram consists
  of three ''permutohedrons'' ( = convex hull of $S^{3}$-orbit of a
point in
  $R^{3}$, where $S^{3}$ acts by permuting coordinates; it resembles
the weight
  system of irreducible representation of $sl(3)$, or $sl(K)$ in the
general $K$
  case, see {\em e.g.} Proposition 11.3 in \cite{Kac}). Their
"highest
  weights" are the points $(6+m,2,0),\/(5+m,2,1),\/(4+m,2,2)$
respectively.

 \section{$K=3$, terms with $\partial Z/\partial T$}\label{oneder}
 They come from the following part:
\begin{equation}\new\begin{array}{c}
\sum_{m,j,k}\frac{\partial^{3}T_{m}}{\partial\Lambda_{ij}\Lambda_{jk}\
Lambda_{ki}}
\frac{\partial Z}{\partial
T_{m}}+\frac{1}{g}\sum_{m,j}\left(2\frac{\partial
g}{\partial\lambda_{i}}
+\right.\\\left.+\frac{\partial
g}{\partial\lambda_{j}}\right)\frac{\partial
T_{m}}{\partial\Lambda_{ij}
\partial\Lambda_{ji}}\frac{\partial Z}{\partial
T_{m}}+\frac{1}{g}\sum_{m,j}
\left(2\frac{\partial
T_{m}}{\partial\lambda_{i}}+\right.\\\left.+\frac{\partial
T_{m}}{\partial\lambda_{j}}\right)
\frac{\partial^{2}g}{\partial\Lambda_{ij}\partial\Lambda_{ji}}\frac{\p
artial Z}
{\partial T_{m}}
\end{array}
      \label{part containing one derivative}
\end{equation}
Differentiating $g$, we obtain :
\begin{equation}   \frac{1}{g}\frac{\partial g}{\partial\lambda_{i}}=
-\frac{1}{3}\sum_{l}\nu_{i}^{3}\nu_{l}\frac{H'(i,l)}{H(i,l)}
                  \label{1st derivative of g}
\end{equation}
where \(H(i,l)=\nu_{i}^{2}+\nu_{i}\nu_{l}+\nu_{l}^{2},H'(i,l)=
2\nu_{l}+\nu_{i} \), and for the second derivatives:
\begin{equation}\new\begin{array}{c}
\sum_{j}\left(2\frac{\partial T_m}{\partial\lambda_i}
+\frac{\partial T_m}{\partial\lambda_j}\right)
\frac{1}{g}\frac{\partial^{2}g}{\partial\Lambda_{ij}\partial\Lambda_{j
i}}
=\\\sum_{k,j}\left(2\frac{\partial T_m}{\partial\lambda_i}+
\frac{\partial T_m}{\partial\lambda_j}\right)
\frac{1}{\mu_{j}^{3}-\mu_{i}^{3}}\left(\frac{\mu_{k}+2\mu_{i}}
{3\mu_{i}^{2}H(i,k)}-\frac{\mu_{k}+2\mu_{j}}{3\mu_{j}^{2}H(j,k)}\right
)+
\\+3\frac{\partial
T_m}{\partial\lambda_i}\left(\frac{2}{27}\mu_{i}^{-6}
+\sum_{k,j}\frac{(\mu_{k}+2\mu_{i})(\mu_{j}+2\mu_{i})}
{(3\mu_{i}^{2})^{2}H(i,k)H(i,j)}\right)
                   \label{2nd derivative of g}
\end{array}
\end{equation}
 Now it is not difficult to write down all the terms. We add the same
expression with $j$
and $k$ exchanged and draw all the terms in the expression (\ref{part
containing one derivative})
(except for the one originating from the last term at the RHS of
(\ref{2nd derivative of g})) through the Newton diagram:

\begin{tabbing}
\=  \\
\>\m{j}\m{\
}\m{2}\m{2}\m{3}\m{3}\m{.}\m{.}\m{.}\m{3}\m{3}\m{3}\m{4}\m{4}\m{
}\m{i}\\
\>\hspace{.20in}\m{
}\m{2}\m{6}\m{8}\m{9}\m{9}\m{.}\m{.}\m{.}\m{9}\m{9}\m{10}\m{12}\m{4}\\

\>\m{
}\m{2}\m{6}\m{6+6}\m{8+6}\m{9+6}\m{9+6}\m{.}\m{.}\m{.}\m{9+6}\m{10+6}\
mr{12+6}\m{20}\m{8}\\
\>\hspace{.20in}\m{
}\m{4}\m{10}\ml{4+12}\m{6+12}\mr{6+12}\m{.}\m{.}\m{.}\m{.}\ml{6+12}\m{
8+12}\mr{16+6}\m{20}\m{4}\\
\>\m{ }\m{
}\m{6}\m{12}\m{18}\m{18}\m{18}\m{.}\m{.}\m{.}\m{.}\m{18}\ml{8+12}\m{12
+16}\mr{12\ }\m{4}\\
\>\hspace{.20in}\m{ }\m{
}\m{6}\m{12}\m{18}\m{.}\m{.}\m{.}\m{.}\m{.}\m{18}\m{6+12}\mr{10+6}\m{1
0}\m{4}\\
\>\m{ }\m{ }\m{
}\m{6}\m{12}\m{18}\m{.}\m{.}\m{.}\m{.}\m{18}\m{.}\m{9+6}\m{9}\m{3}\\
\>\hspace{.20in}\m{ }\m{ }\m{
}\m{.}\m{.}\m{.}\m{.}\m{.}\m{.}\m{.}\m{.}\m{9}\m{3}\\
\>\m{ }\m{ }\m{ }\m{
}\m{.}\m{.}\m{.}\m{.}\m{.}\m{.}\m{.}\m{.}\m{.}\m{3}\\
\>\hspace{.20in}\m{ }\m{ }\m{ }\m{
}\m{.}\m{.}\m{.}\m{.}\m{.}\m{.}\m{.}\m{.}\m{.}\\
\>\m{ }\m{ }\m{ }\m{ }\m{
}\m{.}\m{.}\m{18}\m{18}\m{6+12}\m{9+6}\m{.}\m{.}\\
\>\hspace{.20in}\m{ }\m{ }\m{ }\m{ }\m{
}\m{6}\m{12}\m{18}\m{6+12}\m{9+6}\m{9}\m{.}\\
\>\m{ }\m{ }\m{ }\m{ }\m{ }\m{ }\m{6}\m{10}\m{6+6}\m{8}\m{3}\\
\>\hspace{.20in}\m{ }\m{ }\m{ }\m{ }\m{ }\m{ }\m{4}\m{6}\m{6}\m{2}\\
\>\m{ }\m{ }\m{ }\m{ }\m{ }\m{ }\m{ }\m{2}\m{2}\m{2}\\
\>\\
\>\m{ }\m{ }\m{ }\m{ }\m{ }\m{ }\m{ }\m{ }\m{k}\\
\end{tabbing}
It's easy to see that this expression can be divided by
\(\nu_{j}^{2}+\nu_{j}\nu_{k}
+\nu_{k}^{2} \), the result is given by the following diagram:
\begin{tabbing}
\=   \\
\>\n{j}\n{
}\n{2}\n{2}\n{3}\n{3}\n{.}\n{3}\n{3}\n{3}\n{3}\n{4}\n{4}\n{ }\n{i}\\
\>\hspace{.15in}\n{
}\n{2}\n{4}\n{6}\n{.}\n{.}\n{.}\n{.}\n{6}\n{6}\n{7}\n{8}\\
\>\n{
}\n{2}\n{4}\n{6}\n{.}\n{.}\n{.}\n{.}\n{.}\n{6}\n{7}\n{8}\n{8}\n{4}\\
\>\hspace{.15in}\n{
}\n{2}\n{4}\n{6}\n{.}\n{.}\n{.}\n{.}\n{.}\n{6}\n{7}\n{7}\n{4}\\
\>\n{ }\n{ }\n{2}\n{4}\n{6}\n{.}\n{.}\n{.}\n{.}\n{.}\n{6}\n{6}\n{3}\\
\>\hspace{.15in}\n{ }\n{
}\n{2}\n{.}\n{.}\n{.}\n{.}\n{.}\n{.}\n{.}\n{6}\n{3}\\
\>\n{ }\n{ }\n{ }\n{.}\n{.}\n{.}\n{.}\n{.}\n{.}\n{.}\n{.}\n{3}\\
\>\hspace{.15in}\n{ }\n{ }\n{
}\n{.}\n{.}\n{.}\n{.}\n{.}\n{.}\n{.}\n{.}\\
\>\n{ }\n{ }\n{ }\n{ }\n{.}\n{.}\n{.}\n{.}\n{.}\n{.}\n{.}\\
\>\hspace{.15in}\n{ }\n{ }\n{ }\n{ }\n{2}\n{4}\n{6}\n{6}\n{6}\n{3}\\
\>\n{ }\n{ }\n{ }\n{ }\n{ }\n{2}\n{4}\n{6}\n{6}\n{3}\\
\>\hspace{.15in}\n{ }\n{ }\n{ }\n{ }\n{ }\n{2}\n{4}\n{4}\n{2}\\
\>\n{ }\n{ }\n{ }\n{ }\n{ }\n{ }\n{2}\n{2}\n{2}\\
\>    \\
\>\n{ }\n{ }\n{ }\n{ }\n{ }\n{ }\n{ }\n{k}\\
\end{tabbing}
The Newton diagram of
\((\nu_{i}^{2}+\nu_{i}\nu_{j}+\nu_{j}^{2})(\nu_{i}^{2}+
\nu_{i}\nu_{k}+\nu_{k}^{2})\) is :
\begin{tabbing}
\=\hspace{1.8in}\m{1}\m{1}\m{1}  \\
\>\hspace{1.6in}\m{1}\m{1}\m{1}\\
\>\hspace{1.4in}\m{1}\m{1}\m{1}\\
\end{tabbing}.
It is easy to see that almost the whole polynomial can be divided by
it, except
for some ''boundary'' terms living near the edges and vertices of
Newton
diagram. (''Boundary terms'' means that the ratio of the number of
these terms and the total number of terms goes to zero when
$m\rightarrow\infty$.)
The structure of these terms depends on the remainder of $m$ modulo
3. Consider for
example the case of \(m\equiv2 \bmod 3\). Then the whole expression
is :
\begin{equation}
\new
\begin{array}{c}
-\frac{1}{18}\left\{2\sum_{p=1}^{(m+4)/3}\nu_i^{3p}
\sum_{{a+b=m+9-3p}\atop{a>0,b>0}}\nu_{j}^{a}\nu_{k}^{b}+\right.\\
+\left\{\frac{\nu_{j}}{1+\nu_{j}/\nu_{i}+\nu_{j}^{2}/\nu_{i}^{2}}\sum_
{p=0}^{(m-2)/3}
\nu_{i}^{5+3p}\nu_{k}^{m+3-3p}+(j\leftrightarrow k)\right\}+\\
+\left\{2\frac{\nu_{j}^{2}}{1+\nu_{j}/\nu_{i}+\nu_{j}^{2}/\nu_{i}^{2}}

\sum_{p=0}^{(m-2)/3}\nu_{i}^{4+3p}\nu_{k}^{m+3-3p}+
(j\leftrightarrow k)\right\}+\\
+\frac{\nu_{k}^{4}\nu_{i}^{m+4}\nu_{j}}{1+\nu_{k}/\nu_{i}+\nu_{k}^{2}/
\nu_{i}^{2}}
+\frac{\nu_{j}^{4}\nu_{i}^{m+4}\nu_{k}}{1+\nu_{j}/\nu_{i}+\nu_{j}^{2}/
\nu_{i}^{2}}+\\
+\frac{2\nu_{i}^{m+5}\nu_{j}^{2}\nu_{k}^{2}}{[(1+\nu_{k}/\nu_{i}+\nu_{
k}^{2}/\nu_{i}^{2})
(1+\nu_{j}/\nu_{i}+\nu_{j}^{2}/\nu_{i}^{2})]}+\\
+\frac{4\nu_{i}^{m+5}\nu_{j}\nu_{k}^{3}}{1+\nu_{k}/\nu_{i}+\nu_{k}^{2}
/\nu_{i}^{2}}
+\frac{4\nu_{i}^{m+5}\nu_{j}^{3}\nu_{k}}{1+\nu_{j}/\nu_{i}+\nu_{j}^{2}
/\nu_{i}^{2}}
+\\ \left.
+\frac{\nu_{j}^{4}\nu_{k}\nu_{i}^{m+4}+\nu_{j}\nu_{k}^{4}\nu_{i}^{m+4}
-\hspace{-2pt}
4(\nu_{j}^{2}\nu_{k}^{2}\nu_{i}^{m+5}+\nu_{j}^{3}\nu_{k}^{3}\nu_{i}^{m
+3})}
{[1+\nu_{k}/\nu_{i}+\nu_{k}^{2}/\nu_{i}^{2}]
[1+\nu_{j}/\nu_{i}+\nu_{j}^{2}/\nu_{i}^{2}]} \right\}

                   \label{longequation}
\end{array}
\end{equation}
We must add to this the term \(-\frac{1}{9}{\cal
M}^{-m-5}\sum_{p=o}^{\infty}\sum_{q=0}^{\infty}
({\cal M}^{3p-1}t_{3p+1}+\\\hspace{-2pt}+{\cal M}^{3p}t_{3p+2}-2{\cal
M}^{3p+1}t_{3p+3})
({\cal M}^{3q-1}t_{3q+1}+\hspace{-2pt}{\cal M}^{3q}t_{3q+2}-2{\cal
M}^{3q+1}t_{3q+3})
 \), \\ori\-ginating
from the second term in the expression (\ref{2nd derivative of g}).
In fact it almost
cancels the last term in (\ref{longequation})
, and in the total expression there are no terms with the double sums
both with infinite
upper limit, thought they could appear when expanding
the product of two denominators in Taylor series.

Finally we obtain the following expression for the terms in
coefficient
before $\frac{\partial Z}{\partial T}$, containing the product of two
times:
\begin{equation}\new  \begin{array}{c}
-\frac{1}{9}\{\sum_{p=1}^{(m+7)/3}{\cal M}^{-3p}
\sum_{a+b=m+7-3p}t_{a+1}t_{b+1}+\\+\sum_{q=0}^{\infty}\sum_{p=0}^{(m+1
)/3}
{\cal
M}^{3q-3p-5}t_{3q+1}t_{m+3-3p}+\\+\sum_{q=0}^{\infty}\sum_{p=0}^{(m+1)
/3}
{\cal
M}^{3q-3p-4}t_{3q+2}t_{m+3-3p}-\\-2\sum_{q=0}^{\infty}\sum_{p=0}^{(m+1
)/3}
{\cal M}^{3q-3p-3}t_{3q+3}t_{m+3-3p}+{\cal M}^{-m-6}t_{1}t_{2}\}
\end{array}
\end{equation}
It can be easily compared with the formula (\ref{MAINFORMULA}):the
first double
 sum gives the proper terms in ${\cal W}_{\geq-6}^{(3)}$ excepting
for the terms with $T_{3n}$,
which are cancelled by the corresponding part of the 4th double sum,
whereas the
2nd and the 3d and some part of the 4th double sum are responsible
for
$T{\cal W}_{\geq-3}^{(2)}$ terms in (\ref{MAINFORMULA}),
and the term with $t_{1}t_{2}$ corresponds to
$\frac{\partial}{\partial T}{\cal W}_{-3}.$
The terms with less than two times in front of $\frac{\partial
Z}{\partial T}$ are the
same as in $N=1$ case \cite{KMMMZ91b}.

In fact we do not need to calculate the terms which do not contain
time derivatives of $Z$.
Actually it is easy to calculate $Z$ up to the three loops. The
result is:
\begin{equation}\new\begin{array}{c}
Z_{3
loops}[T]=1+\frac{1}{3}(T_{1}^{2}T_{2}+\frac{1}{3}T_{4})+\frac{1}{18}T
_{1}^{4}
T_{2}^{2}+\frac{13}{27}T_{1}^{2}T_{2}T_{4}+\\+\frac{5}{27}T_{1}^{3}T_{
5}+
\frac{7}{27}T_{1}T_{7}-\frac{2}{27}T_{2}^{4}+\frac{13}{162}T_{4}^{2}
\end{array}
\end{equation}
(The simple way to obtain this expression is as follows. Calculate
first
$\tau_{0L}+\tau_{2L}$ by the direct computation or using Hirota
equations
and ${\cal L}_{-1}$ -constraint,
then calculate 3-loop terms depending on $T_{1}$ using ${\cal
L}_{-1}$-constraint
and prove by using $(D_{1}^{4}+3D_{2}^{2})\tau\cdot\tau=0$ that there
is no term
$T_{2}^{2}T_{4}$ on the 3-loop level. Therefore,
$\tau_{3L}=\tau_{2L}+f(T_{1})+\beta T_{4}^{2}
+\gamma T_{8}$. Find $\frac{1}{16}\beta+\frac{1}{8}\gamma$ using
$N=1$ calculations.
Then, $\gamma=0$ because of absence of 3-loop fat graph with one
boundary:
corresponding Riemann surface would have
$dimH^{0}-dimH^{1}+dimH^{2}=1-3+1=-1$,
that is impossible since Euler characteristics of orientable
2-manifold must be
even.) 3-loop $Z$ is sufficient to determine the terms without
derivatives in
${\cal W}$.

The terms with $\alpha$ can be easily calculated and compared with
the corresponding terms
in (\ref{MAINFORMULA}).

 \section{Appendix}\label{appendix}
 Here we'll consider the coefficient in front of \(\frac{\partial
Z}{\partial T}\) and
show how the cancellation of non-desirable terms could arise in
general $K$ case.

 This coefficient is given by the following expression:
 \begin{equation}
 \new
 \begin{array}{c}
 \left(\frac{\partial^{K}
T_{m}}{\partial\Lambda_{tr}^{K}}\right)_{ii}+
 \sum_{j_{1}\ldots j_{K-2}}
 \left(2\frac{\partial\ln g}{\partial\lambda_{i}}+
 \sum_{p=1}^{K-2}\frac{\partial\ln g}{\partial\lambda_{j_{p}}}
 \right)\frac{\partial^{K-1}T_{m}}
 {\partial\Lambda_{ij_{1}}\ldots\partial\Lambda_{j_{K-2}i}}+\ldots
 \end{array}
 \label{A1}
 \end{equation}
 As in the case $K=3$, the bulk of the terms in the numerator will
fill some
 $K-1$--dimensional polygon in space where Newton diagrams live, and
their are
 some ''boundary'' terms of codimensions 1,2, etc., living near the
edges of the
 corresponding codimensions. The denominator is equal to
\[\prod_{0\leq p<q\leq K-1}
 h_{K-1}(\nu_{j_{q}},\nu_{j_{p}}), \; j_{0}:=i.\] We'll now restrict
ourselves to
 consideration of the terms with codimensions 0 and 1, which
correspond to the
 bulk of the terms in Ward identities.

 The expression ~(\ref{A1}) is equal to:
 \begin{equation}
 \new
 \begin{array}{c}
 \frac{(-1)^{K}}{K}\sum_{j_{1}\ldots
j_{K-1}}\{(\nu_{i}\nu_{j_{1}}\ldots
 \nu_{j_{K-1}})^{K}
 \frac{\left|\begin{array}{ccccc}1&\cdot&\cdot&\cdot&1\\\cdot&\cdot&\c
dot&\cdot&
 \cdot\\
 \nu_{i}^{(K-1)K+m}&\cdot&\cdot&\cdot&\nu_{j_{K-1}}^{(K-1)K+m}\end{arr
ay}\right|}
 {\left|\begin{array}{ccccc}1&\cdot&\cdot&\cdot&1\\\cdot&\cdot&\cdot&\
cdot&\cdot\\
 \nu_{i}^{(K-1)K}&\cdot&\cdot&\cdot&\nu_{j_{K-1}}^{(K-1)K}\end{array}\
right|}+
 \\+\frac{
\frac{\nu_{i}^{K}H'(i,j_{K-1})}{H(i,j_{K-1})}+\hspace{-1.5pt}\sum_{0\l
eq q<K-1}\hspace{-4pt}
\frac{\nu_{j_{q}}^{K}H'(j_{q},j_{K-1})}{H(j_{q},j_{K-1})}}{(\nu_{i}\nu
_{j_{1}}\ldots
\nu_{j_{K-2}})^{-K}K\nu_{j_{K-1}}^{-1}} \/ \frac{\left| \!
\begin{array}{ccccc}
1&\cdot&\cdot&\cdot&1\\
\cdot&\cdot&\cdot&\cdot&\cdot\\\nu_{i}^{(K-2)K+m}&\cdot&\cdot&\cdot&
\nu_{j_{K-2}}^{(K-2)K+m}\end{array} \!
\right|}{\left|\begin{array}{ccccc}1&\cdot&\cdot&
\cdot&1\\\cdot&\cdot&\cdot&\cdot&\cdot\\\nu_{i}^{(K-2)K}&\cdot&\cdot&
\cdot&\nu_{j_{K-2}}^{(K-2)K}\end{array}\right|}+\\
 +\mbox{(boundary terms of codimension $>$1)} \}
 \end{array}
 \label{A2}
 \end{equation}
 Here we use the notations:\[\nu_{j}=\mu_{j}^{-1}, \:
 H(i,j)=h_{K-1}(\nu_{i},\nu_{j}), \:
 H'(i,j)=(K-1)\nu_{j_{K-1}}^{K-2}+\ldots+\nu_{i}^{K-2}\].
 \newtheorem{lemma}{Lemma}
 \begin{lemma}
 \begin{equation}
 \new
 \begin{array}{c}
 \left|\begin{array}{ccccc}1&\cdot&\cdot&\cdot&1\\
 \nu_{i}^{K}&\cdot&\cdot&\cdot&\nu_{j_{\alpha-1}}^{K}\\
 \cdot&\cdot&\cdot&\cdot&\cdot\\
 \nu_{i}^{(\alpha-1)K+m}&\cdot&\cdot&\cdot&
 \nu_{j_{\alpha-1}}^{(\alpha-1)K+m}\\\end{array}\right|=\\=\Delta(\nu)

 \{\nu_{i}^{(\alpha-1)(K-1)+m}\prod_{0<r<s\leq\alpha-1}H(j_{r},j_{s})+
\\+
 h_{1}(\nu_{j_{1}},\ldots,\nu_{j_{\alpha-1}})\nu_{i}^{(\alpha-1)(K-1)+
m-1}
 \prod_{0<r<s\leq\alpha-1}H(j_{r},j_{s})+\ldots\}
 \end{array}
 \label{formula}
 \end{equation}
 \end{lemma}
\newtheorem{remark}{Remark}
\begin{remark}
{\em Such a form of writing this expression can give a wrong
impression that this
expression can be divided by
\(\prod_{0<r<s\leq\alpha-1}H(j_{r},j_{s})\).

\begin{picture}(170,100)
\put(150,50){Fig.1}
\thicklines
\put(30,20){\line(1,0){90}}\put(30,20){\line(-1,1){20}}
\put(120,20){\line(1,1){20}}\put(10,40){\line(1,1){50}}
\put(60,90){\line(1,0){30}}\put(140,40){\line(-1,1){50}}
\thinlines
\put(124,24){\line(-1,0){98}}
\put(128,28){\line(-1,0){106}}
\put(132,32){\line(-1,0){114}}
\put(136,36){\line(-1,0){122}}
\put(140,40){\line(-1,0){130}}
\label{picture1}
\end{picture}

In fact this formula reproduces all the terms but those in the
vicinity of the
''bottom'' of the corresponding Newton diagram---these terms are
represented by
dashed area in the figure~\ref{picture1}, which represents $K=3$
case. These terms
are determined unambiguously by the symmetry of the Newton diagram
(in the figure~\ref{picture1}, this is the symmetry with respect to
rotation on
\(\pi/3\)).}
\end{remark}
[{\em Proof.}]Note that the structure of terms near the $i$-th  top
does not
 depend on the class of $m$ modulo $K$, and we can assume $m=Kn, n\in
Z$ without
 loss of generality. Then it is easy to see \cite{Macdonald} that
 \begin{equation}
 \left|\begin{array}{ccccc}1&\cdot&\cdot&\cdot&1\\
                     \cdot&\cdot&\cdot&\cdot&\cdot\\

\nu_{i}^{(\alpha-1+n)K}&\cdot&\cdot&\cdot&\nu_{j_{\alpha-1}}^{(\alpha-
1+n)K}
  \end{array}\right|=\Delta(\nu^{K})h_{n}(\nu^{K})
  \end{equation}
  Thus the structure of the terms not in the vicinity of bottom is
the same as
  in the expression
  \begin{equation}
  \prod_{0\leq p<q\leq\alpha-1}(\nu_{j_{p}}-\nu_{j_{q}})
  \prod_{0<r<s\leq\alpha-1}H(j_{r},j_{s})
  \prod_{0<r\leq\alpha-1}H(i,j_{r})h_{n}(\nu_{k})
  \end{equation}
  It's easy to see considering the Newton diagram that multiplying
  \(h_{n}(\nu^{k})\) by \(\prod_{0<r\leq\alpha-1}H(i,j_{r})\) we
obtain
  \(h_{m+(\alpha-1)(K-1)}(\nu)\) minus some boundary terms located
near the
  bottom (see fig.2).

\begin{picture}(300,110)
\put(125,1){Fig.2}
\put(60,70){\circle*{4}}
\multiput(45,40)(30,0){2}{\circle*{4}}
\multiput(30,10)(30,0){3}{\circle*{4}}
\put(120,50){\vector(1,0){60}}

\put(240,90){\circle*{2}}               \put(240,70){\circle*{4}}
\multiput(235,80)(10,0){2}{\circle*{2}}
\multiput(230,70)(10,0){3}{\circle*{2}}
\multiput(225,60)(10,0){4}{\circle*{2}}
\multiput(225,40)(30,0){2}{\circle*{4}}
\multiput(220,50)(10,0){5}{\circle*{2}}
\multiput(215,40)(10,0){6}{\circle*{2}}
\multiput(210,30)(10,0){7}{\circle*{2}}
\multiput(210,10)(30,0){3}{\circle*{4}}
\multiput(205,20)(10,0){8}{\circle*{2}}

\multiput(240,90)(15,-30){3}{\line(1,-2){10}}
\multiput(240,90)(-15,-30){3}{\line(-1,-2){10}}
\multiput(255,60)(15,-30){2}{\line(-1,-2){10}}
\multiput(225,60)(-15,-30){2}{\line(1,-2){10}}
\multiput(250,70)(15,-30){2}{\line(-1,-2){10}}
\multiput(230,70)(-15,-30){2}{\line(1,-2){10}}
\multiput(240,30)(-5,10){2}{\line(-1,-2){10}}
\multiput(240,30)(5,10){2}{\line(1,-2){10}}
\label{picture2}
\end{picture}

Lemma is proved.
\newtheorem{assumption}{Assumption}
\begin{assumption}
{\em We assume that the boundary terms on the bottom edges of
codimension 2 match
the terms near the bottom facet of the Newton diagram in such a way
that
the whole expression after symmetrisation over
\(\nu_{j_{1}},\ldots,\)\\ \(\nu_{j_{K-1}}\)
is dividable by} \(\prod_{0<p<q\leq\alpha-1}H(j_{p},j_{q})\), {\em
the result being some polynomial}
\(F(\nu_{i},\nu_{j_{1}},\ldots,\nu_{j_{K-1}})\)
{\em, such that for every $p$ the bulk of the terms in $F$,
containing \(\nu_{i}^{p}\),
has the same coefficient ({\em i.e.} the multiplicities of the
corresponding points
in the Newton diagram are equal). (cf. the case $K=3$).}
\end{assumption}
In fact the strongest requirement in this assumption is that the
polynomial can
be divided without remainder by the above-mentioned expression. This
is equivalent
 to the statement that the terms in Ward identity can be expressed
through times. This
 is a miracle we can not explain and prove in general case.
 The second requirement (about the form of $F$) seems to be much more
natural,but
 we also cannot prove it.

Then we remain with the expression:
\begin{equation}
\new
\begin{array}{c}
\frac{(\nu_{i}\nu_{j_{1}}\ldots\nu_{j_{K-1}})^{K}}{K}
\frac{h_{m+(K-1)^{2}}(\nu_{1},\ldots,\nu_{j_{K-1}})}{\prod_{0<p\leq
K-1}H(j_{p},i)}+\\
+\frac{(\nu_{i}\nu_{j_{1}}\ldots\nu_{j_{K-2}})^{K}}{K^{2}}
\left(\nu_{i}^{K}\nu_{j_{K-1}}\frac{H'(i,j_{K-1})}{H(i,j_{K-1})}+
\sum_{p=0}^{K-1}\nu_{j_{p}}^{K}\nu_{j_{K-1}}\frac{H'(j_{p},j_{K-1})}{H
(j_{p},j_{K-1})}\right)\times\\
\times\frac{h_{m+(K-2)(K-1)}(\nu_{1},\ldots,\nu_{j_{K-2}})}{\prod_{0<p
\leq K-2}H(j_{p},i)}
+\mbox{boundary terms}
\end{array}
\end{equation}
\label{remain with}
Of course, $F$ cannot be divided by \(\prod_{p}H(i,j_{p})\) for
general $K$.
But now we'll construct the boundary terms which could ensure such a
division
and really find these terms in the second summand in (\ref{A1}).
It is obviously that $F$ can be divided by this expression
with remainder, situated near the bottom. But we know that there are
no boundary
terms of codimension 1 near the bottom (we know all boundary terms of
codimension 1, see the equations (\ref{A1}),(\ref{A2})). So we need a
procedure
to divide this polynomial with no remainder near the bottom. To this
end we
consider the system of parallelotops, obtained from the right side of
the figure
 2 shifting it by the integer lattice vectors in the directions
 perpendicular to the $i$-th axis,see fig.3.

\begin{picture}(200,100)
\thicklines
\put(85,90){\line(-1,-1){60}}\put(85,90){\line(1,-1){60}}
\put(25,30){\line(1,0){120}}
\thinlines
\multiput(55,80)(10,0){7}{\line(-1,-1){10}}
\multiput(55,80)(10,0){7}{\line(1,-1){10}}
\multiput(55,60)(10,0){7}{\line(1,1){10}}
\multiput(55,60)(10,0){7}{\line(-1,1){10}}

\multiput(41,65)(10,0){10}{\line(-1,-1){10}}
\multiput(41,65)(10,0){10}{\line(1,-1){10}}
\multiput(41,45)(10,0){10}{\line(1,1){10}}
\multiput(41,45)(10,0){10}{\line(-1,1){10}}

\multiput(25,50)(10,0){13}{\line(-1,-1){10}}
\multiput(25,50)(10,0){13}{\line(1,-1){10}}
\multiput(25,30)(10,0){13}{\line(1,1){10}}
\multiput(25,30)(10,0){13}{\line(-1,1){10}}

\put(65,10){Fig.3}
\label{picture3}
\end{picture}

This system is equal to $F$ up to the following
boundary terms:
\begin{equation}
\new
\begin{array}{c}
\frac{(-\nu_{i}\nu_{j_{1}}\ldots\nu_{j_{K-2}})^{K}}{K^{2}
\prod_{0<p<K}H(i,j_{p})}(\nu_{j_{K-1}}h_{m+(K-1)K}
(\nu_{i},\ldots,\nu_{j_{K-2}})+2\nu_{j_{K-1}}^{2}h_{m+(K-1)K-1}+\\
+\ldots+(K-1)\nu_{j_{K-1}}^{K-1}h_{m+(K-1)^{2}-1})+\hspace{-1.5pt}
(\mbox{cyclic permutations of
\(j_{1},\ldots,j_{K-1}\)})\hspace{-1.5pt}+\\+(\mbox{boundary
terms of codimension $>1$})
\end{array}
\label{boundary}
\end{equation}
To see this, one must calculate carefully the multiplicities of
points
on fig.3 near the bottom and compare them with the multiplicities
of the corresponding terms in $F$, which can be easily calculated
(see
(\ref{formula}) and remark after it). The boundary terms of
(\ref{boundary}) arise
from the ''boundary'' parallelotops on fig.3.

Consider now the second term in the expression \ref{remain with}.
Note that if we change in this term
\(\nu_{j_{q}}^{K}\nu_{j_{K-1}}H'(j_{q},j_{K-1})
\prod_{p\geq 0,p\neq q,\/p\leq K-1}H(j_{K-1},j_{p})\)
 by \\ \(\nu_{i}^{K}\nu_{j_{K-1}}H'(i,j_{K-1})\prod_{p>0,p\leq
K-2}H(j_{K-1},j_{p})\)
it in fact is equivalent to a shift of the Newton diagram in
\((i\cdots j_{K-2})\)
plane, and so changes nothing but 2-codimensional boundary terms. So
the whole
expression is equal modulo these terms to the sum of $K$ terms each
of them
being equal to:
\begin{equation}
\frac{(-\nu_{i}\ldots\nu_{j_{K-2}})^{K}\nu_{i}^{K}\nu_{j_{K-1}}H'(i,j_
{K-1})
h_{m+(K-1)(K-2)}(\nu_{i},\ldots,\nu_{j_{K-2}})}{K^{2}\prod_{0<p\leq
K-1}
H(i,j_{p})}
\end{equation}
It is obvious that $K-1$ of this terms will give just appropriate
boundary
terms (cf (\ref{boundary})) when we exchange $j_{K-1}$ with
$j_{1},\ldots,j_{K-2}$. The remaining term
is equal modulo boundary terms to:
\begin{equation}
\frac{\{\sum_{p=1}^{K-1}\nu_{j_{K-1}}^{p}\nu_{i}^{K-p}-\hspace{-.3pt}(
K-1)\nu_{j_{K-1}}^{K}\}
h_{m}(\nu_{i}^{K},\ldots,\nu_{j_{K-2}})}{K^{K-1}(-\nu_{i}\ldots\nu_{j_
{K-2}})^{-K}}
\sum_{q=0}^{\infty}\left(\frac{\nu_{j_{K-1}}}{\nu_{i}}\right)^{Kq}
\label{remaining term}
\end{equation}
Compare it with the first term in (\ref{A1}), for which we have just
found that the remainder
cancels with the boundary terms and so can divide it by the
denominator, obtaining
$(-)^{K}\frac{(\nu_{i}\ldots\nu_{j_{K-1}})^{K}}{K^{K-1}}
h_{m}(\nu_{i}^{K},\nu_{j},\ldots,\nu_{j_{K-1}})$. We see that terms
in
$h_{m}(\ldots)$, containing one $\nu_{j>0}^{K}$ and all other
$\nu_{j>0}$
with the powers $\neq0\bmod K$ are exactly cancelled by the last term
in $\{\ldots\}$
in (\ref{remaining term}), while other terms in $\{\ldots\}$
constitute terms like
$\nu_{i}^{pK-r}t_{r}h_{pK}(t)$.

So we see that at least the bulk of the coefficients in front of
$\partial Z/\partial T$ have
the structure like that in the expression:
\begin{equation}
\new
\begin{array}{c}
(-1)^{K}tr \: \epsilon{\cal M}^{-K^{2}}\left(\sum_{p\geq-K+1}
{\cal M}^{-pK}{\cal W}_{pK}^{(K)}\right.
+\\+K\sum_{{p+a\geq-K+2}\atop{a\geq1}}\sum_{r=1}^{K-1}{\cal
M}^{-pK+r}
(aK+r)T_{aK+r}{\cal W}_{(a+p)K}^{(K-1)}-\\
-K(K-1)\sum_{{p+a\geq-K+2}\atop{a\geq1, p\leq-K}}{\cal
M}^{-pK}aKT_{aK}{\cal W}
_{(a+p)K}^{(K-1)}+\ldots\left.\right)
\end{array}
\end{equation}
where ${\cal W}$ are normalized as in \cite{FKN91b}.

Of course, these considerations are very far from the real proof. In
fact, new effect
appears even at $K=4$. Namely, the condition of cancellation of
denominators
of the type \(\frac{1}{\nu_{j}^{K-1}+\ldots+\nu_{k}^{K-1}}\) is
nontrivial now
not only for the terms which do not contain $\alpha$, but also for
those which contain
it. This condition seems to be very restrictive.
  \section{Acknowledgments}
I thank A.Morozov for stating the problem and discussions. I am also
indepted to
S.Kharchev, A.Marshakov, A.Mironov, S.Piunikhin, M.Vasiliev and
B.Voronov for
discussions.

         \end{document}